\documentclass[times, 10pt,twocolumn]{article} 
\usepackage{latex8}
\usepackage{graphicx}
\usepackage{times}

\begin{document}

\title{COMODI: On the Graphical User Interface}

\author
{Zsolt I. L\'az\'ar\\
Faculty of Physics\\ 
Babe\c{s}-Bolyai University\\
Str. M. Kog\~{a}lniceanu Nr. 1, RO 400084\\
Cluj-Napoca, Romania\\
zlazar@phys.ubbcluj.ro,
\and
Andreea Fanea,
Drago\c{s} Petra\c{s}cu,\\
Vladiela Ciobotariu-Boer
and Bazil P\^{a}rv\\
Faculty of Mathematics and Computer Science\\ 
Babe\c{s}-Bolyai University\\
Str. M. Kog\~{a}lniceanu Nr. 1, RO 400084\\
Cluj-Napoca, Romania\\
\{afanea,petrascu,vladi,bparv\}@cs.ubbcluj.ro
}

\maketitle
\thispagestyle{empty}

\begin{abstract}
We propose a series of features for the graphical user interface (GUI) of the COmputational MOdule Integrator (COMODI) \cite{Synasc05a}\cite{COMODI}. In view of the special requirements that a COMODI type of framework for scientific computing imposes and inspiring from existing solutions that provide advanced graphical visual programming environments, we identify those elements and associated behaviors that will have to find their way into the first release of COMODI. 
\end{abstract}

\section{Introduction}

The COmputational MODule Integrator (COMODI)\cite{COMODI} is an interdisciplinary collaboration for addressing the problem of programming crisis in computational sciences \cite{Post}. In \cite{Synasc05a}, \cite{Lazar05cse}  and \cite{Lazar04a} we have pointed out that there are at least three major problems plaguing computational software development and usage. Computational scientists tend to spend much of their time on writing code already written by others or trying to adapt existing code to local needs. On global scale this represents a costly \textit{lack of efficiency} in exploiting human resources.  Generally, \textit{low quality} ``home made" code is the only alternative to the anguish of searching, adapting and learning third party software; a process that further undermines the trust in external code. Finally, the \textit{irreproducibility} of computer experiments by peers  contributes again to the inefficiency of computational research by wasting the important advantage of virtuality over real experiments. We suggest that shifting towards a reuse oriented paradigm is necessary. In the same  papers it is argued that a new software tool, language or standard is not sufficient. One has to explicitely aim at a large scale movement. However, the movement can only evolve around an OpenSource software solution that has to offer both on the short and long run sufficient benefits so that the computational community with deeply rooted programming traditions would consider it as an alternative, and contribute actively to its development. COMODI has the sole reponsibility of igniting the process and then letting the events follow their natural course driven by the need and skill of the whole community. Therefore, it should not be a radically futuristic solution but rather a sturdy bridge between the old and new paradigm. 
During the ignition process, due to the lack of involvement of the community the task is especially challenging. The needs that are best to target are difficult to identify and because of limited human resources the evolutionary trial and error approach that keeps OpenSource even though inefficient yet healthy is not admissible. Hence, the requirement analysis is the core issue of our investigation. In \cite{Synasc05a} we point out that the complete solution consists in two distinct parts. On one side there are tools used for converting regular code developed by component authors into a COMODI component, called the \emph{developer side}, while on the other, there is a GUI-based graphical environment with underlying layers responsible for binding together components into arbitrarily complex computational projects displayed as in figure \ref{project}. We term the latter as the \emph{user side}. The two sides are clearly distinguishable by the tools that are used, the type of human activity that they presume and the required programming skills. Present paper focuses on the user side exploring its most prominent component, the GUI. 
As shown in \cite{Synasc05a}, \cite{Lazar05cse} and \cite{Lazar04a} the idiosyncrasy of COMODI is the special support provided to component authors. On the user side, GUIs for visual programming can look back to a relatively long evolution history. Several applications endowed with advanced features are in use today. Coming up with something that would assure a leading edge is more than a challenge and is off the point. However, the constituting premises of COMODI entail a series of special requirements set for the user interface. These, while do not make it a direct competitor in this segment, allow for using present solutions as a valuable source of inspiration in its design process. Below we summarize a few important solutions worthwhile to consider.

\section{Visual programming in scientific computing}

Though visual programming (VP) is not a widely spread phenomenon in computational science, the two have already made contact in a few areas \cite{vp1}. Visualization of scientific data is clearly a leader in this respect. \textit{AVS} (Advanced Visualization System) is one of the dominant commercial visualization packages available today. It provides modules and a user interface which allows these modules to be connected together into a flow chart. The compatibility of the interfaces is checked automatically. Input data is processed sequentially by the modules and usually an image is produced. AVS thus implements a data-flow paradigm of scientific visualization. However, the developers of new AVS modules have an involved procedure to follow. \textit{OpenDX} (Open Data eXplorer) is an OpenSource variant of AVS with similar features. It comes with hundreds of built-in specialized functions. One can contribute with new components by following an API. The new component will become available after recompiling the application. 
Outside data visualization commercial solutions dominate the market. \textit{Simulink} is a software package for modeling, simulating, and analyzing dynamic systems. It provides a GUI for building models as block diagrams, using drag-and-drop mouse operations. One can also customize and create ones own blocks. \textit{LabVIEW} is a ``program development application''. It provides a graphical programming language, as opposed to a text-based language, used to create programs in a block diagram form for data acquisition and presentation. IRIS Explorer and the associated numerical and visualization libraries is another attempt for a tool to increase productivity.

In the realm of open-source visual programming there has been much less progress, but that is mainly due to the fact that the attempts are rather new. Based on a language interoperability tool called Babel, a couple of projects have been started, each producing, for the moment, a rudimentary GUI. SciRun2 is a more mature computational framework that recently has turned to Babel to increase its connectivity \cite{vp2}.

In spite of the availability of the above and other high quality tools a solution that would deeply penetrate the community is still lacking. The reason is either being of limited scope, e.g. only visualization, or technically too challenging for component authors with natural sciences or engineering background, or because of not supporting the traditional ways of coding, or too expensive, or not supported by the majority, or all the above. COMODI is trying to fill in the above gaps. 
 
\section{The COMODI GUI}

As suggested in the introduction the user side GUI is not the part that would bring about the envisioned break-through. Even if quickly acquiring the support of a large user and developer community, the GUI will still have to go through a long maturing process before it can close up on similar commercial VP environments. However, it can significantly limit the impact of COMODI if the design of the very first release doesn't follow such clear principles as the \textit{zero effort threshold} commandment on the component development procedure \cite{Synasc05a}. 
Below we formulate a few requirements that are essential for COMODI to be appealing:
\begin{itemize}
 \item the learning curve for exploiting the capabilities of the system should be smooth: The interdependence of the accessible features and the level of expertise required for using them should be free of abrupt thresholds. A careful design of default behaviors and wizards is essential; 
 \item it has to offer all the benefits of high level programming yet should keep low-level control at easy reach: The user should be able to set the level of abstraction that is wanted. Visualizing connectors, breakpoints and other ``administrative" components should be possible but optional;
 \item self-explanatory and easy to use: the inspection of components' interfaces should not require more than a couple of clicks. Tooltips, status bar texts and alike for the documentation of the component, its ports, and the ports' parameters should be ubiquitous; 
 \item highly-customizable: components should look similar enough so that using them would always be obvious but allow for customization both by the author of the component and the user. Similarly to controlling fonts in a browser, users should be able to override certain author settings. Many will prefer to see features similar to existing VP environments. 
\end{itemize}

We consider that sequentiality, an idea that is strongly imprinted into scientific software developers' vision, must also be suggestively represented by the relative arrangement of components in a project graph and the location of ports on a component. In the early stages of the maturing process of COMODI, the programming interfaces of components will likely be designed at will, without respecting any standards. Hence, during the assembling process the user will have to manually match provides and uses ports, and check the compatibility of the connected ports at the level of each argument. Therefore, the inspection of all ports and parameters therein will have to be assisted by GUI elements made as handy as possible.

In perspective, we expect that a reuse oriented community will enforce standards which not only that will speed up considerably the manual assembling process but will open the way to automatic interface matching (search $\rightarrow$ (down)load $\rightarrow$ paste $\rightarrow$ configure) managed by AI modules.

\section{Representation of components and ports}

\begin{figure}[!htbp]
\centerline{\includegraphics[width=6.5cm]{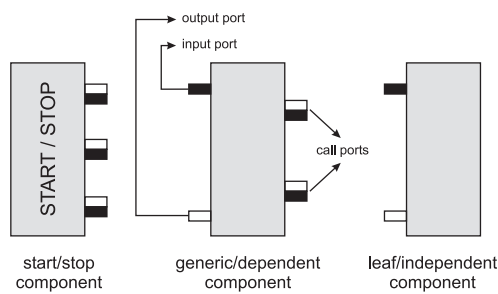}}
\caption{Component classification}
\label{port_types}
\end{figure}

For the first release of COMODI, components are regarded as the lowest granularity units of code, namely, functions and procedures.   On the left-hand side of a component we have its \emph{provide} ports while on the right its \emph{uses} or \emph{call} ports (see figure \ref{port_types}). We shall term the two as the provide and uses side, respectively. Data flows in/out via the black/white ports. They represent entry and exit points of regular function calls. Provide ports are connection points to the parent component located one level higher in the call tree. Conversely, uses ports connect the component to its children, one level deeper in the call hierarchy. This representation is fairly intuitive as it maps almost directly to the source code wherein the body of a function contains lines with the invocation of other functions. 

As a general rule the diversification of component representation should be fully supported in order to assure guides for the eye, that is, a better overview of large project graphs that extend over several screens manageable only via scrolling and zooming. The difference in the components' functionalities should be reflected in their appearance. Therefore we suggest that all uses ports should be displayed. Many components will still seem alike but this simple rule will confer certain variety to the outlook of components. Statistically speaking, there will be a correlation between the weight of a component in terms of offered services and its geometric size. In perspective, the recommended solution is a \emph{skin repository} referenced by components and downloaded by the framework upon request.

\begin{figure}[!htbp]
\centerline{\includegraphics[width=6.5cm]{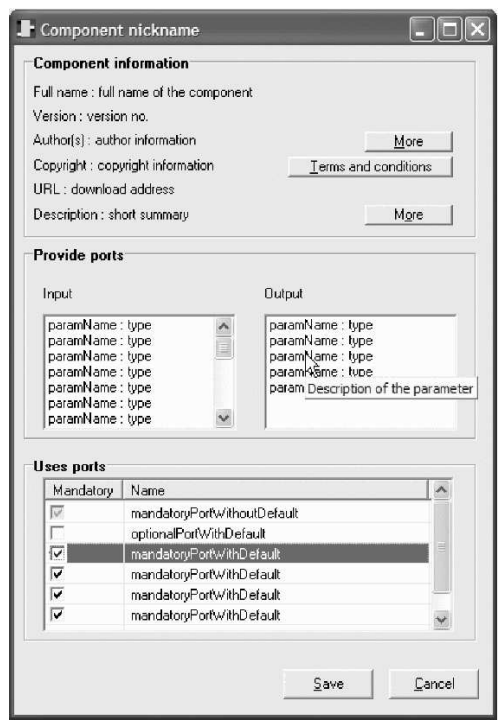}}
\caption{Graphical user interface for inspecting and setting component properties}
\label{component_information}
\end{figure}

In order to manage connections the user should be provided with interfaces that make the inspection of all component, ports and parameter properties straightforward. Figure \ref{component_information} presents an interface containing most relevant information pertaining to a component. The window can be accessed by double-clicking the component in the project graph.

\begin{figure}[!htbp]
\centerline{\includegraphics[width=6.5cm]{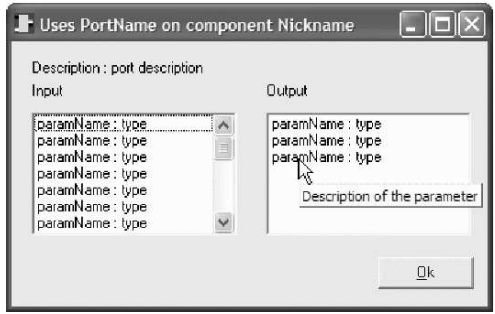}}
\caption{Graphical user interface for inspecting port properties}
\label{port_information}
\end{figure}

The \textit{Uses ports} section displays the names of the ports as referred to by the author, leaving the actual identity of the child components to the user to decide. Double clicking one of the uses ports from the list would lead to another information window dedicated to that particular port, as shown in figure \ref{port_information}. Direct click on a uses port in the project graph opens up this very same window. 
Future versions will allow linking of complex data types in the parameter lists to corresponding documentation windows. 

\subsection{Mandatory and optional ports.} 

In the reuse oriented future, components should not simply expose some ``dangling bonds'' or, on the contrary, be completely autonomous. They should allow for as many uses ports as possible but in the same time provide from within the deployed package a default connection for all these ports. For instance, a molecular dynamics component should allow the user to re-define any of the  two- and three-body interaction functions between chemical elements. On the other hand, a component handling $k$-body interactions of $n$ chemical elements would be completely useless if $n!/(n-k)!k!$ different  functions would need to be wired up before using it. Figure \ref{component_information} demonstrates a possible alternative for handling this issue. For most, ideally all uses ports a default connection will be defined, i.e., an exported or private function within the deployed component. Using the checkboxes the user selects those ports that are to be connected to user-defined external components. Those that are unchecked use the default internal references. Inactive checkboxes represent the case for ports that do not have an internal default connection therefore are always displayed on the right side of the component and user provided connection is mandatory.
Another possibility is to use external components as default connections within the boundaries of a composed component.


\section{Representation of a project}

\begin{figure}[!htbp]
\centerline{\includegraphics[width=6.5cm]{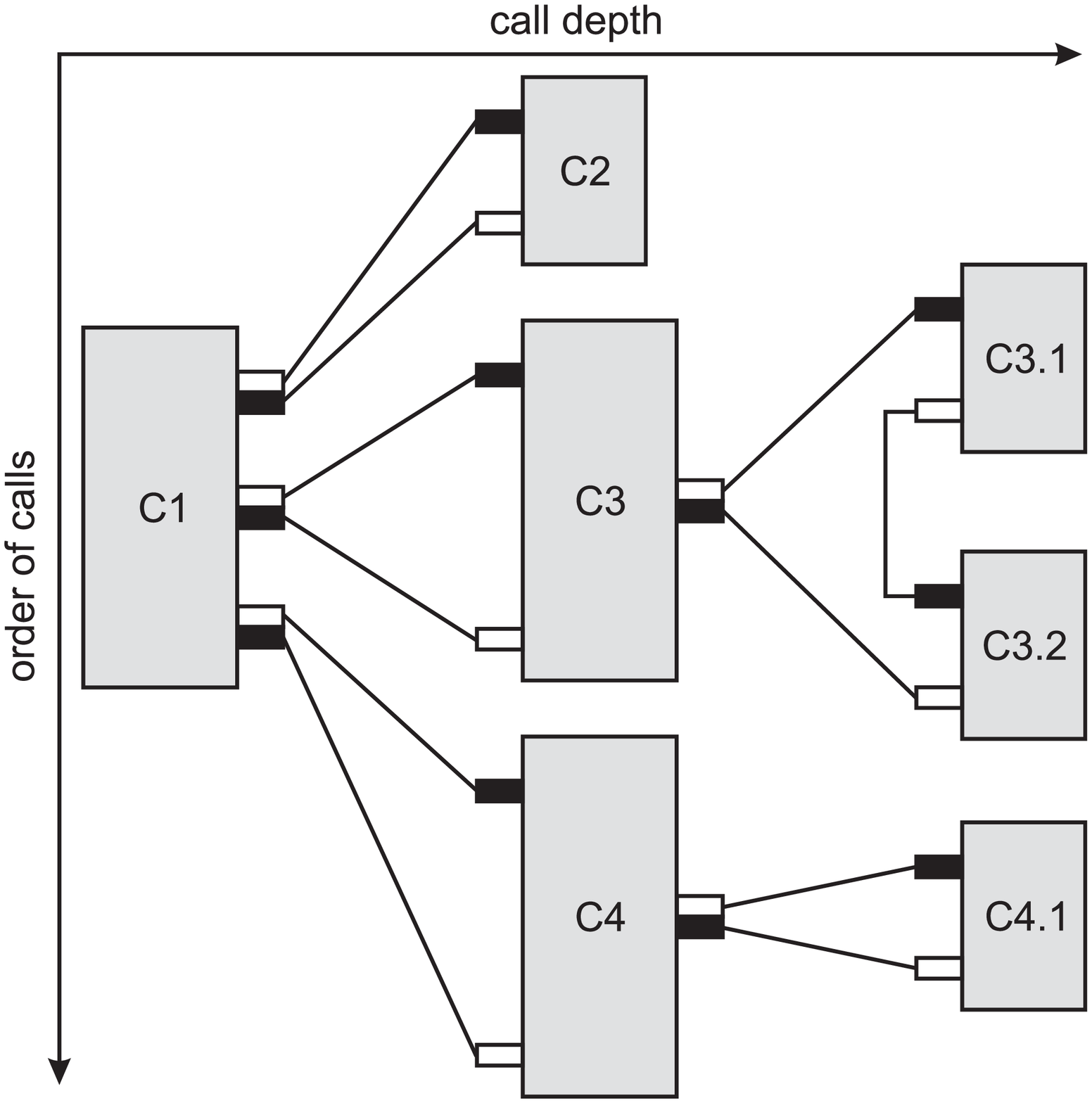}}
\caption{Recommended representation of a project in the framework COMODI. 
The components reach execution state starting from left to right, bottom to 
top. In this case the order is: framework (C1), C2, C3, C3.1, C3.2, C4, C4.1}
\label{project}
\end{figure}

A \textit{project} is a graph of interconnected components, as shown in figure \ref{project}.  
\textit{Connection} means data flow via function interfaces from 
the output of one component to the input of another \cite{Lazar04b}.
We can identify two different types of communication: a horizontal \emph{client-server} type of a communication entailing a change in the call depth and a vertical \emph{pipe{\&}filter} type entailing data transfer between siblings. C3 -- C3.2, exemplifying a client-server communication is the direct representation of the pull model for module communication. It consists in a function calling another, which returns data to the former. This is basically the only model for communication in low-level languages such as C or Fortran.  On the other hand, vertical calls (C3.1 -- C3.2) imply the piping of an output into the input ports of another component. This push model, is not directly supported by low-level languages. However, as we shall show below it can be transformed into equivalent client-server communication. Disregarding this aspect the structure is tree-like. At this stage, branching and loops are concepts that are unknown to the project. All decisions pertaining to the order of child calls are made runtime and inside the components. Hence the order of calls, the way it is suggested in figure \ref{project}, is not strict. It is rather meant to suggest the position of the corresponding call point in the source code of the component.  In perspective, the extraction of non-tree elements (branches, loops) from inside the components out into the flow diagram is conceivable with the help of  specialized connector components.

In order to allow smoothless transition between GUI versions or even completely different environments -- just as different windows managers can be plugged into the same X Windows system -- COMODI should come with a basic DTD that describes the representation independent state of the system, e.g., topology of the project, execution state, etc. All GUIs come with their own DTDs that are extensions of the representation independent one. In case a GUI encounters a project or a component with an incomprehensible descriptor file, relying on the services of the framework, it will find the ``greatest common denominator" in the DTD inheritance hierarchy available online, transforms the descriptor file accordingly and then proceed with the rendering.

\section{Connectors}

Connectors are simple components mediating dataflow between ``real'' components. This indirection can be due to reasons such as the necessity for satisfying certain syntactical requirements of component wiring, splitting up outgoing dataflow and sharing it between child components, merging incoming dataflow, broadcasting the same data to several components, etc. For a review on connectors see \cite{Mayer02} and ch. 10 in \cite{Szyperski02}. 

\begin{figure}[!htbp]
\centerline{\includegraphics[width=6.5cm]{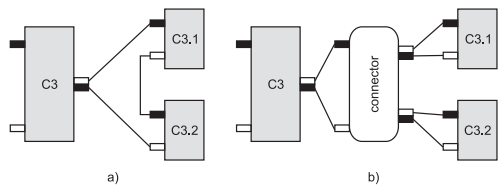}}
\caption{Translating between pipe\&filter and client-server communication via a connector.}
\label{connector}
\end{figure}

Here we only want to single out one particular type of connector. Above we noted that pipe\&filter communication is an abstraction that can be realized by transforming it into two client-server calls via a connector. Figure \ref{connector}a and \ref{connector}b illustrates this transformation. As an additional feature the GUI can display the connector for those who are keen of seeing a perfect tree hierarchy or hide it if preferred. While this connector can be managed automatically, most of the others will need specialized user interfaces for manual configuration. 

\section{Conclusions and outlook}

In spite of a great deal of similarities with existing visual programming environments, the special scope of the COMODI project endows the user framework's GUI with features that are not common in other solutions. The suggested graphical elements and behavior are intuitively in harmony with the low-level programming paradigm. 
Once released to the public domain, a fast development of the GUI is expected. Parallel to the development of the above presented GUI several others are expected to emerge targeting restricted groups of computational scientists. Therefore, during the design process a special emphasis should be laid on defining proper interlayer communication interfaces. 

\section*{Acknowledgment}
This work is supported by the National University Research Council of Romania with grant no. 27687/14.03.2005.

\end{document}